\documentclass{elsart}
\usepackage{epsfig}
\begin{document}
\begin{frontmatter}

\title{The critical properties  of the agent-based model with 
       environmental-economic interactions}

\author{      Zolt\'an Kuscsik, Denis Horv\'ath and Martin Gmitra}  
             
 \address{    Department of Theoretical, Physics and Astrophysics, 
              University of P.J. \v{S}af\'arik,                                  
              Park Angelinum 9, 040 01 Ko\v{s}ice, Slovak Republic                                     
          }






\begin{abstract}
The steady-state and nonequilibrium properties of the 
model of environmen\-tal-econo\-mic interac\-tions are studied. 
The interacting heterogeneous agents are simulated on the platform 
of the emission dynamics of cellular automaton. The model  
 possess the discontinuous transition between 
the safe and catastrophic ecology. Right at the critical line, 
the broad-scale power-law distributions of emission rates have 
been identified. Their relationship to Zipf's law and models of 
self-organized criticality is discussed. 
\end{abstract}
\end{frontmatter}

\section{Introduction}\label{Int}

On the onset of global ecological crisis 
 questions related to regulation technologies and 
safe handling of environmental resources become urgent. 
The equilibrium between the economic growth and environmental 
variables is an actual problem of every government.
The environmental science 
must deal with the task how to design 
 robust sensorial 
nets and control systems to manage 
the balance between the immediate 
economic needs and the long-time 
perspectives of mankind. There are several instruments for control and decrease the local 
and global pollution rate: energy taxation,  pollution rights and introduction of cleaner 
technologies. Unfortunately, 
direct and naive application of these approaches does not result relevant emission reductions.  
Enforcing pollution rights is difficult and has high administrative costs~\cite{Pearson1995,Speck1999}. 
As an example can serve the multi-country control model of international coordination of emission charges and 
pollution trading~\cite{Ploeg1991}. As shown in Ref.~\cite{Ploeg1991} pollution trading may result a non-cooperative 
equilibrium with high pollution rates. It is often argued that if the only tool of policy is 
the energy taxation it may cause the loss of the competitiveness between firms. Moreover, the 
tax rate which is be able to stabilize the emissions would be too high to be economically 
and politically acceptable. Because of energy taxation and network effects in society, the firms tend 
to delay adoption to cleaner technologies~\cite{Carraro1997,Eftichios2005}. All these unexpected 
emergent regulatory effects stimulated our interest about the problem 
of dynamics of emissions that can be seen as a consequence of the competitive interaction of the environmental 
and economical (acquisition) agents.

A key principle we applied to define a robust environmental 
policy is the feedback mechanism. 
In general, the nonlinear 
feedback is one of the key up-bottom mechanism that enables to keep control 
over the dynamics of complex systems. 
The mechanism has been exploited in many different systems. 
The problem discussed in~Ref.\cite{Melby2000} is 
focused to the feedback localization of the parametric boundary 
between periodicity and chaos generated by a logistic map. 
Stabilization of dynamical regimes via feedback appears in the simulations of 
investments~Ref.\cite{Holyst2000}. In~Ref.\cite{Youss1998,Bradford1990} 
model has been suggested, where the speculative market expectations 
are linked via the positive feedback to the prevailing price trend.  
As stated in~Ref.\cite{Kadanoff}, the steady {\em self-organized critical} 
(SOC) regime~Ref.\cite{Bak88} implies the operation of inherent feedback 
mechanism that ensures a steady state marginally stable against 
disturbances. It is important here to note that {\em criticality} is accompanied by the absence of  
characteristic scale of the {\em power-law distributions}  of
 avalanches. According~Ref.\cite{Sornette} the nonlinear feedback allows to transform 
the models of the "unstable" phase transitions (i.e. classic models) 
to SOC models. In the works devoted to self-adjusted Monte Carlo 
methods~Ref.\cite{Tomita2001,HorvathGmitra2004}, the feedback which 
drives the physical system to a critical regime  is expressed in terms 
of estimated statistical averages. Additionally, catastrophic environmental collapses 
in consumption/pollution feedback models were 
described in~\cite{Clarke1994,Tsur1998}. In this paper we present 
feedback control distributed over the population of autonomous sensing 
mobile agents that control the behavior of 
factory agents. 

As usual, the thinking about distributed control is motivated 
by the expected robustness against accidental failures. Nevertheless,
most of agent systems face to the general problem of agency 
that is the imperfect incentives~\cite{Axtell} to satisfy the global goal function of 
 system. 
Our motivation is to show, that this property of distributed agent system may drive 
the environment to a critical state with extremal emission ratio. 
The idea of walking sensorial agents in our model is 
very similar to the idea of distributed pollution 
sensing by pigeon bloggers proposed by 
de Costa~Ref.\cite{deCosta2006}, where pigeon agents are 
equipped by communication chips and sensors 
for carbon monoxide. Similar idea had been reported 
as a concept of randomly walking sensors for detection 
of dynamically changing gradient sources~\cite{Dhariwal}.

The plan of the paper is as it follows. In the next 
section the multi-agent ecological model is introduced. 
In section sec.\ref{numres} the results of numerical study 
are discussed. Finally the conclusions are presented. 

\section{The model}
 
 The first point we consider
 is the dynamics of diffusive emissions 
 of factory agents. 
 The base substance of model is a square lattice with dimension $L \times L$ which represents
 the physical space for emission spreading.
 The factories are located on random lattice sites 
 and the spatial distribution of their emissions is directed 
 by discrete rule of diffusion.
 For this purpose, we implement a cellular automaton (CA)
 based on integer rules. Our suggestion for computation of diffusion is partially
 inspired by the lattice gas models 
 of pollutants~Ref.\cite{Marin2000}.
  However, we preferred formulation where 
 the concentration field on lattice sites varies 
 via the recursive rule 
 \begin{equation}
  \forall {\bf R}\in \{ L\times L \} \,,  \qquad   m(t+1,{\bf R})  = 
 m(t,{\bf R})   +  I_{\rm nn}(t,{\bf R}) +  
 I_{\rm F}(t,{\bf R})  
\label{dif2}
\end{equation}
 that replaces in one 
 time step 
 $t$ the $m(t,{\bf R})$ "integer pieces" of emission
 at the position ${\bf R}\equiv [R_x,R_y]$.   The 
 emission inflow  caused by the factory agents $I_{\rm F}$  
 is discussed in below~Eq.(\ref{IFeq2})
 The emission current between the nearest neighbor 
 cells 
 \begin{equation}\label{INN}
 I_{\rm nn}(t,{\bf R})
  \equiv   -
  4 [ m(t,{\bf R})/ 5]_{\rm int}
  +  \sum_{{\bf r}\in {\rm nn}({\bf R})} \,
  [ \, m(t,{\bf r})/5 ]_{\rm int}\,\,,
  \label{cure2}
 \end{equation}
 where the brackets $[\ldots]_{\rm int}$  denote 
 the integer part of argument and  $\sum_{  {\rm nn}({\bf R}) }$ 
 refers to the four nearest neighboring sites of 
 ${\bf R}$. Eq.(\ref{INN}) 
 preserves the total emission for transitions between the lattice sites .
 Other words, the natural purification of emission is absent, 
   emissions are exclusively 
 depleted by the outflow 
 through opened lattice bounds. For $m(t,{\bf R})\gg 1$ 
 the mass equidistribution of emissions becomes satisfied 
  with $1/\sqrt{m}$ order.  Under these conditions, 
 $I_{\rm nn}(t,{\bf R})$  acts as 2d Laplace of $m$,  
 and thus CA rules describe the stylized diffusive transport. 
  
 The local concentration of emissions is permanently monitored  
 by $N_{\rm S}$ number of {\em sensing agents}.
The sensing agents are 
 random walkers  with the lattice positions 
 ${\bf S}(t,j)  \equiv [S_x(t,j)$, $S_y(t,j)]$, $j=1,2,\ldots$, $N_{\rm S}$. 
 The lattice position of the $j$th walker  
 is given by 
\begin{eqnarray}\label{walker}
{ S_{ x} }(t+1,j) &  =  &  [ S_{ x} (t,j) + s_{ x}(t,j)]_{\bmod L}\,, \\
\nonumber
{ S_{ y} }(t+1,j) & =   & [ S_{ y} (t,j) + s_{ y}(t,j)]_{\bmod L}\,;
\end{eqnarray}
where the shift ${\bf s}(t,j) \equiv [s_x(t,j), s_y(t,j)]$ 
is chosen randomly from the set of four unit vectors 
$\{ (0,1),$  $(0,-1),$  $(1,0),$  $(-1,0)\}$. Eq.(\ref{walker}) preserves the periodic  
 boundary conditions for ${\bf S}(t,j)$.
Moreover, we checked  that the statistical 
dependencies  studied in this paper do not change remarkably for closed boundary conditions.

The source of emissions are the
\emph{factory agents} localized at random positions 
${\bf F}(t,k)\in \{L\times L\}$, $k=1,2,\ldots, $ $N_{\rm F}$ . 
They are characterized 
by an integer value of 
emission production $n(t,k)$. 
 The condition of sensorial feedback system 
is that over a
predetermined local pollution 
threshold  $m_{\rm c}$ every sensorial agent with
$m(t,{\bf S}(t,j)) > m_{\rm c}$ sends a feedback request to  
the nearest~(in the Euclidean sense) factory 
${\bf F}(t,k_{\rm near}^{(j)})$  
\begin{equation}
\min_{k=1,2,\ldots, N_{\rm F}} 
\,\|\, {\bf S}(t,j)  - {\bf F}(t,k)\, \| 
= \|\,{\bf S}(t,j) - {\bf F}(t,k_{\rm near}^{(j)}) \|\,.  
\end{equation}
The concentration threshold 
 $m_{\rm c}$ 
is analogous to that in~Ref.\cite{Cammann1992} where the 
feedback operation is mediated by the chemical 
sensors with specific thresholds for different 
pollutants. Any request of sensorial agent instantly decreases 
the local emission production $n(t,k)$ of the $k$th factory
by a constant additive factor $\Delta>0$. In the absence of  
feedback signal, to satisfy economic needs,
the factory increases its production automatically.
Both alternatives are described by the equation   
\begin{eqnarray} 
\forall k, \qquad
n(t+1,k)= \left\{ 
\begin{array}{lll}
n(t,k) 
+   \Delta   &    &   
\mbox{\small $ 
\begin{array}{l}
\mbox{when no signal}  \\
\mbox{from any ${\bf S}(t,j)$ agent } \\ 
\mbox{is received}
\end{array}$} 
\\
&&
\\
\max\{n(t, k)-\Delta;0\}    &  & 
\mbox{\small $
\begin{array}{l}
 \mbox{feedback signal} \\
\mbox{received}
\end{array}
$} 
\end{array}
\right.. 
\end{eqnarray}
The eventuality $n(t,k) - \Delta \le 0 $ (zero emission means no production) 
yields the bankruptcy of ${\bf F}(t,k)$ and
the appearance of a new factory 
 at random position  with initial production $n(t+1,k) =0$.
The zero initial emission assure that new factory may start its production
only when it survives at least one time period without feedback request.
The total number of factories is conserved by the death-birth rules.
Since the production strategies of new generation of factories does not change, 
the implicit assumption here is that system is locked to the inferior 
technology~\cite{Arthur1989}.

What remains unspecified in Eq.(\ref{dif2}) is the emission inflow caused by the
 ${\bf F}$ type of agents
\begin{equation}
I_{\rm F}(t,{\bf R})  =  
n(t,k)\, \delta_{ {\bf R},   {\bf F}(t,k)   }\,.
\label{IFeq2}
\end{equation}
It is worth nothing that all the agent systems in common 
with CA subsystem are updated synchronously. 
The asynchronous update rules can be also of interest, 
but we dismiss them from our considerations.

After the completion of the agent-based model, its relationship to the 
standard sandpile model of SOC~\cite{Bak88} should be noted. 
The level of emissions is confronted  with a threshold which is similar to the height (slope) threshold in 
 SOC models. In sandpile models if the height of a pile exceeds a given threshold, 
a "signal" is transmitted to the central site and to its neighboring 
sites. The abundant grains are then toppled from 
the center. Spatial expansion of a sandpile avalanche  corresponds to the emission spreading around a
 factory agent. The power-law distribution of avalanches 
in sandpile model is a consequence of the local conservation 
of sand grains. The property coincide with the conservation 
of emissions, but contradicts to the non-local conservation 
of factory agents. The opened boundary conditions for emissions 
are also borrowed from the sandpile model.
\section{Simulation results}\label{numres}
The simulations have been performed for lattices 
$L=20, 30, 40, 50$. To avoid 
the influence of transients, the initial $10^5$ updates 
per site were discarded; about 
$\tau = 5 \times 10^7$ updates have been used to collect 
information about the steady-state for the fixed parameters 
$m_{\rm c}=5 \gg \Delta =1$. In further we limit ourselves 
to the systems with $\rho_{\rm F}=N_{\rm F}/L^2=1/100$. 
In the world with large number of sensors $N_{\rm S}$ the emissions 
fluctuate around $m_{\rm c}$. This oversaturated 
population of the sensors prevents from unbearable 
contamination but yields a non-efficient economy. 
Much different is the situation for $N_{\rm S}$ small, 
where the steady-state regime become unstable since 
the rate of emissions diverges in time. The  
dynamics corresponds to a catastrophic regime. 
The preliminary numerical work on steady-state
statistics yields the conclusion that $m(t,{\bf R})$ is remarkably lowered
at the corners of $L\times L$ lattice world with opened bounds.
Such inhomogeneity induces secondary effect of the
accumulation of factories at boundary regions.
As $L$ increases, the overall contribution of corners decays faster
than the contribution of edges. In this context, the important
is the aspect of finite-size scaling carried out in the next section.

\begin{figure}
\centerline{
\includegraphics[width=0.8\textwidth]{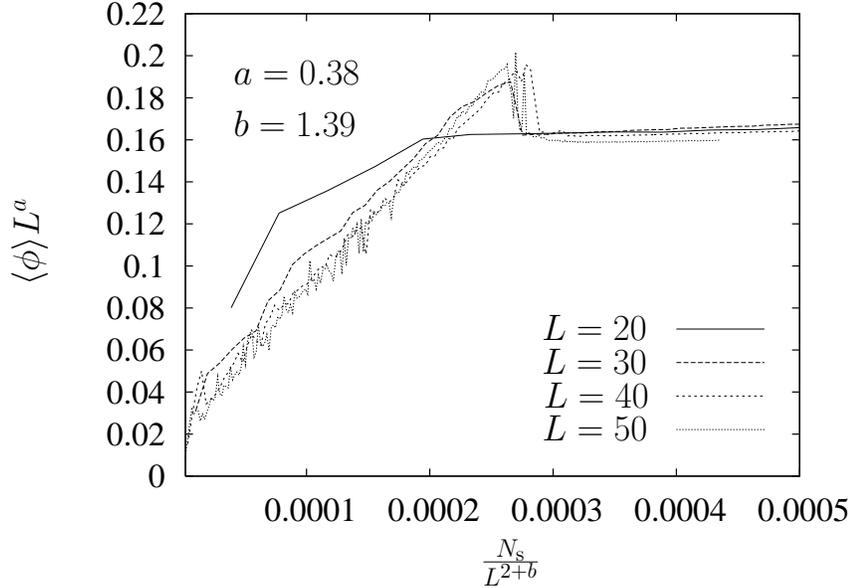}}
\caption{
Finite size scaling of the order parameter $\langle \phi \rangle$ as 
a function of sensorial agent density $N_{\rm S}/L^2$ scaled by $L^{-b}$ factor. The simulation was performed for different $L$ with fixed density 
of factories $\rho_{\rm F} = 0.01$. }
\label{fig1}
\end{figure}

The outcome of our 
simulation experience is the decision
to accumulate the statistics of those 
factories that have received 
even one feedback request during the step $t$. 
Thus, we define the following order parameter
\begin{equation} 
\langle \phi \rangle = \frac{1}{\tau} \sum_{t=0}^{\tau-1} \phi(t)\,,
\quad
\phi(t) = \frac{1}{N_{\rm F}}\sum_{k=1}^{N_{\rm F}} \sigma(t,k) \,,
\label{Eqsigma1}
\end{equation}
 where $\langle \phi \rangle $ is the temporal average calculated over $\tau$ epochs.
 If source  ${\bf F}(t,k)$ received a 
feedback signal $\sigma(t,k)=1$ and $\sigma(t,k)=0$ 
otherwise. Fig.\ref{fig1} shows the $N_{\rm S}$ dependencies of 
$\langle \phi \rangle$ that attain abrupt jump interpreted as a discontinuous transition 
form the safe (nondestructive) to the destructive economy.
Moreover, we found that $N_{\rm S}$ dependencies of $\langle \phi \rangle$
collapse onto the single finite-size scaling form 
\begin{equation}
\langle \phi \rangle =  
L^{-a}\, f( \theta \,)\,, \qquad  \theta=  N_{\rm S}/ L^{2+b}
\label{critical1}
\end{equation}
with exponents $a=0.38$, $b=1.39$. 
The algebraic combination 
$\theta_{\rm crit}= N_{\rm S}/ L^{2+b}$,   where 
the scaling function $f(\ldots)$ attains the jump 
$f(\theta_{\rm crit})$ and 
$\langle \phi \rangle$ 
tends to the $L^{-a}$ asymptotics, is interpreted 
as a critical point. For 
given ${\rho_{\rm F}}$ we have estimated $\theta_{\rm crit} \simeq 2.8 \times 10^{-4}$. 
For sufficiently small 
lattices ($L\leq20$) the scaling violation is indicated.

\begin{figure}\label{scale_all1}
 \centerline{  \includegraphics[width=0.8\textwidth]{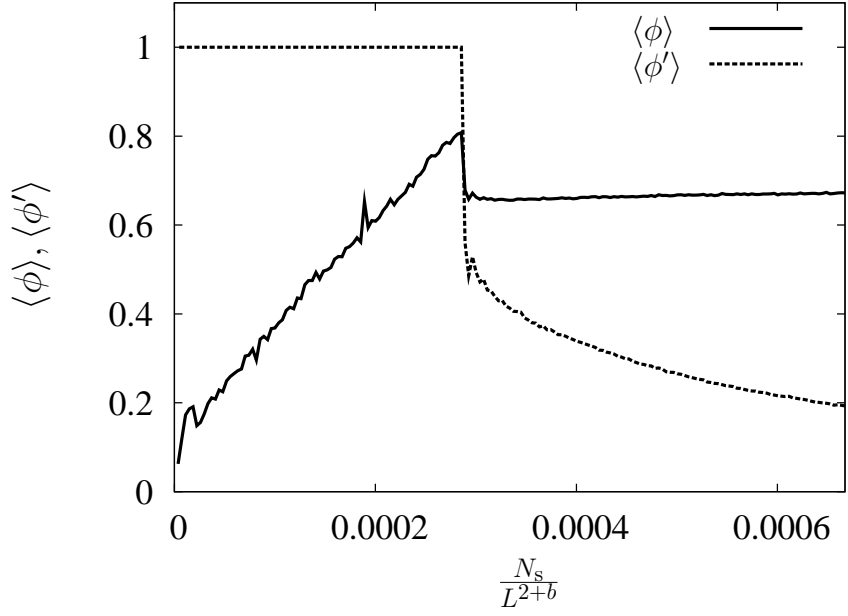}}
   \caption{ 
    The order parameters as function of sensorial agent density $N_{\rm S}/L^2$ scaled
    by $L^{-b}$ factor for $L=40$. The other parameters as in Fig.1.}   
   \label{fig2}
 \end{figure}

The problem with sensorial system manifests itself when 
$\theta< \theta_{\rm crit}$. In this regime a sensor multiply 
affects the same nearest factory and lowers its 
emission production, while other hidden sources, 
maybe more influential and dangerous polluters, 
could left without feedback requests. 
Hereafter, this process is referred as {\em spatial screening}. 
The screening emerges as a result of {\em co-evolution}
and {\em self-organization} in multi-agent system. 
The statistical significance of the screening 
is clarified in Fig.\ref{fig2} where the auxiliary order parameter 
\begin{equation}
\langle \phi' \rangle = \frac{1}{\tau} \sum_{t=0}^{\tau-1} \phi(t)\,,
\quad
\phi'(t) = \frac{1}{N_{\rm S}}\sum _{j=1}^{N_{\rm S}} \sigma'(t,j)\,,
\label{auxil1}
\end{equation}
is plotted against $\theta$. The $\sigma'(t,j)=1$ when the request comes from the $j$th sensorial agent, $\sigma'(t,j)=0$ otherwise. 
At the critical point $\theta=\theta_{\rm crit}$ both the order
parameters attain discontinuous jump. The ratio 
$N_{\rm S}/N_{\rm F}$ reflects the {\em screening} efficiency
which at the critical point reads
$N_{\rm S} / N_{\rm F} =  
\theta_{\rm crit} L^b /\rho_{\rm F} = 
\theta_{\rm crit} \rho_{\rm F}^{-(b+2)/2} N_{\rm F}^{b/2}$. 
The proportionality $N_{\rm S}/N_{\rm F}  \sim N_{\rm F}^{b/2}$ 
reveals the interpretation of the index $b$.
The difference $\langle \phi' \rangle -\langle \phi \rangle$ 
reflects the impact of the screening as well as the most pronounced situation at $\theta\leq \theta_{\rm crit}$.

\begin{figure}
  \label{difag}
\centerline{ \includegraphics[width=0.8\textwidth]{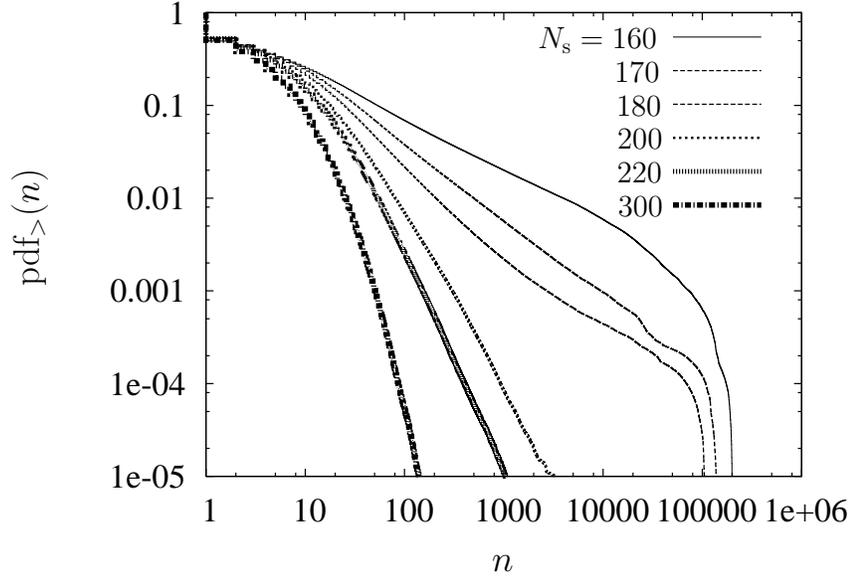}}
 \caption{  The log-log plot of the cumulative distribution 
  of emissions $n(t,k)$ for $L=50$. The boundary
  power-law regime ($N_{\rm S}=160$) is the most pronounced. 
 Thereupon for $N_{\rm S}<160$ the dynamics has no steady
state and steady-state  distributions do not exist.}
 \label{fig3}
 \end{figure}

\begin{figure}\label{scale}
\centerline{   \includegraphics[width=0.8\textwidth]{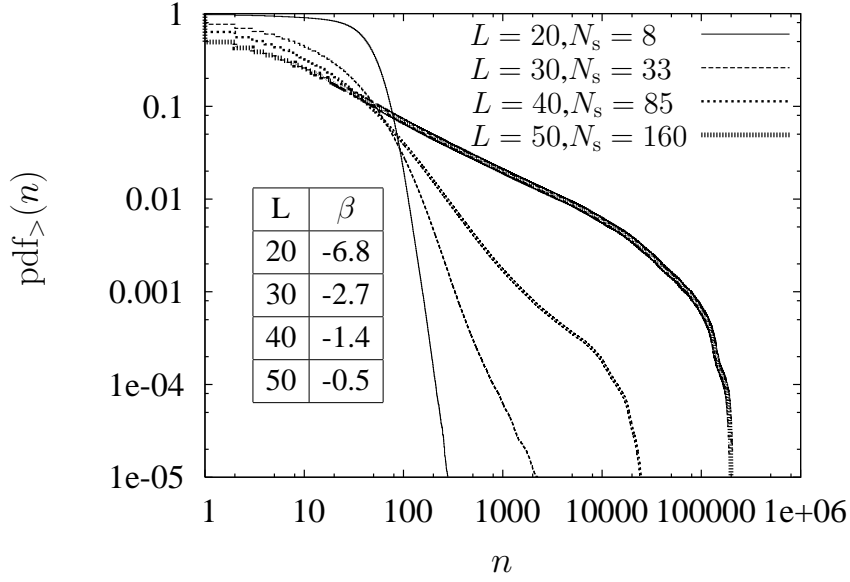}}
   \caption{ 
     The cumulative pdf's of the production rates $n(t,k)$ obtained 
    for different  $L$  and $N_{\rm S}(L)=\theta_{\rm crit} L^{b+2}$
    The inset table lists the effective 
    exponents $\beta$.}  
   \label{fig4}
   \end{figure}

The vicinity of criticality is corroborated by the broad-scale 
statistics of production rates (see Fig.\ref{fig3}).  
We see that the most pronounced 
power-law dependence belongs 
to $\theta\simeq \theta_{\rm crit}$. 
In the economic models Ref.\cite{Ploeg1991}, the emission rates 
are usually related to the overall 
production of goods with a proportionality factor 
called emission-output ratio. Standard and
widely accepted assumption is that this ratio is fixed 
and irrespective to the investments in clean technology. 
With such idealization, the critical fluctuations 
of $n(t,k)$ can be assigned to the empirical Zipf's law.  
As we see from Fig.\ref{fig4}, the cumulative probability distribution 
function $\mbox{pdf}_{>}(n)$ corresponding 
to ($N_{\rm S}, L$)- parametric domain have been 
characterized by the effective exponents within the 
broad range $\beta \simeq 0.5-6.8$. It means that feedback mechanism 
we proposed is not sufficient to fix the realistic exponents of the 
Zipf's law (according Ref.\cite{Fujiwara2004} the empirical exponents 
are $\beta=0.84$;\,\, $0.995$). 

Since the place for new factory is chosen randomly, the snapshots 
showing the spatial distributions of ${\bf F}$ agents evoke the images 
of the short-range ordered molecular patterns. From this 
perspective the system of factories have features of 2d molecular 
liquid or "vapor". Within such viewpoint the "vapor" of ${\bf F}(t,k)$ 
agents prepared for $\theta>\theta_{\rm crit}$ belongs 
to the safe economy, whereas stronger correlated "liquid condensate" 
($\theta< \theta_{\rm crit}$) corresponds to the catastrophic regime. 
In further, these arguments are applied in favor of scenario of discontinuous transition. 
\begin{figure}
\centerline{\includegraphics[width=0.7\textwidth]{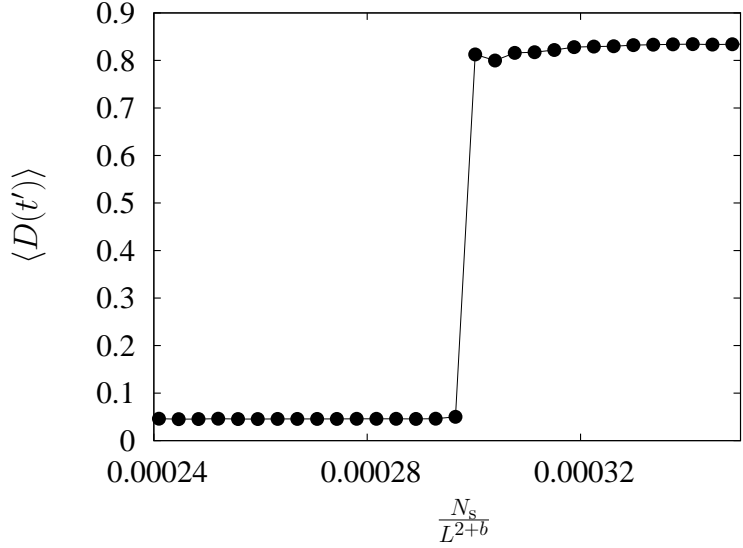}}
\caption{ 
 The application of DS technique for $L=40$, $t'=250$. 
 The averaged inter-replica distance  
 $D(t')$ [see Eq.(\ref{demspr2})] is plotted  
 versus $\theta = N_{\rm S}/L^{2+b}$.  
 The position 
 of the critical point coincides  with 
 Fig.\ref{fig1} within the statistical error. 
}
\label{fig5}
\end{figure}
\begin{figure}
\centerline{\includegraphics[width=0.7\textwidth]{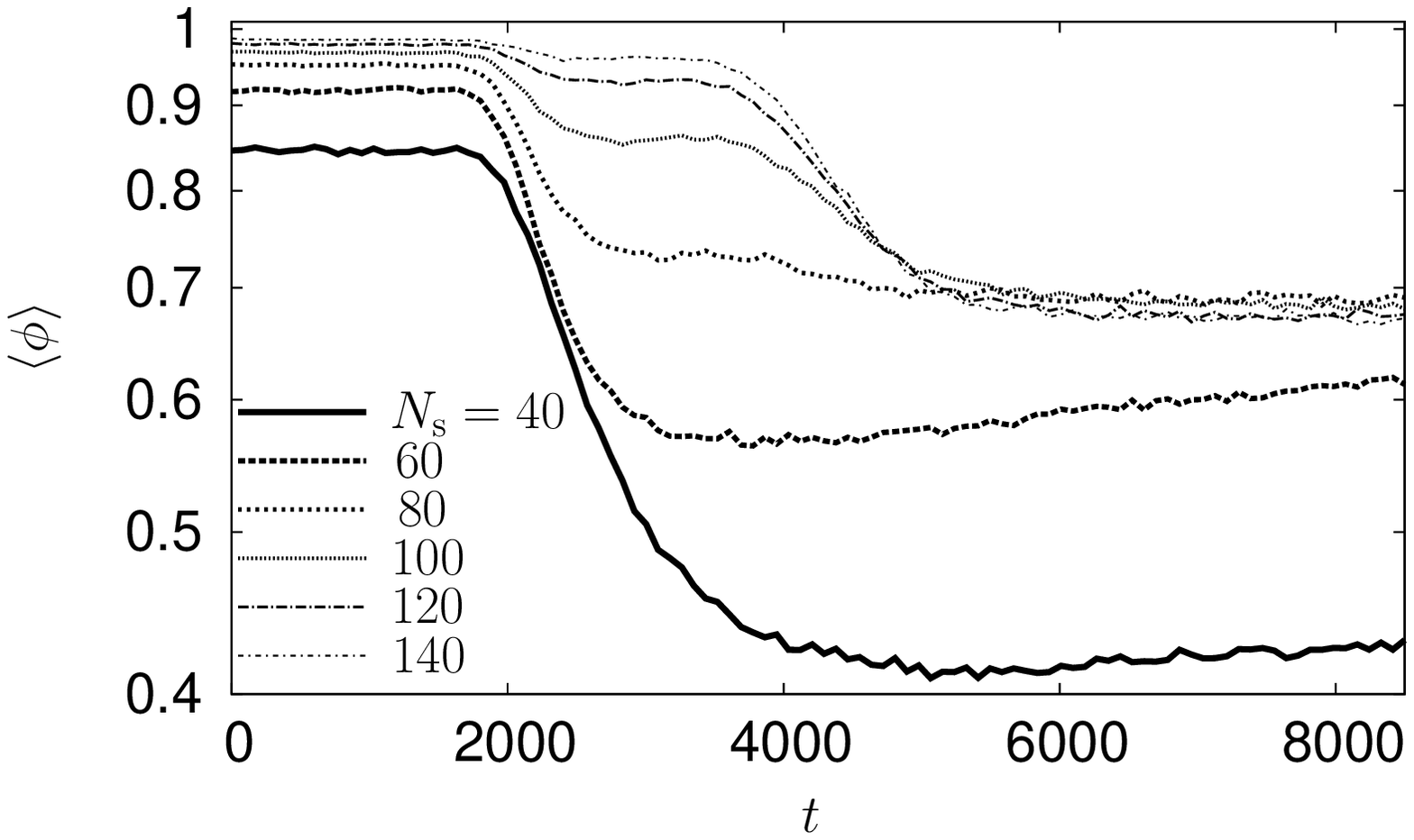}}
 \caption{ The differences between nonequilibrium relaxation 
below and above the critical number (i.e. minimum number) 
of sensors sufficient to prevent from dangerous 
pollution is $N_{\rm S}\simeq 80$ (for $L=40$). 
The interval ($2000<t<4000$) approximately 
delimits the fast CA purification 
through the opened bounds. 
The steady-state 
becomes formed above 
$t\simeq 7000$ 
(above the crossing point).} 
\label{fig6}
\end{figure}
The transition concept is examined in further 
(in a manner that is independent of $\langle \phi\rangle$) 
via damage spreading~(DS) method Ref.\cite{Kauffman69}. 
The DS is based on the comparative analysis of 
simultaneously evolving replicas. Let us consider the 
pair of the systems I,II with equal number of agents. 
At certain time $t$ the copy II is replicated from 
I and, consequently, single grain of the emission (damage) 
is added to its randomly 
chosen cell ${\bf R}'$: 
\begin{equation}
\forall {\bf R}\,,\,\,\, t'=0\,, 
\qquad
m^{\rm (II)}(0,{\bf R})=  
m^{\rm (I)}(t,{\bf R}) +  \delta_{{\bf R},{\bf R}'}\,\,.
\label{Ini2}
\end{equation}
The subsequent $t'$ time steps 
are executed 
with the identical sequences of random numbers 
for updating of the both of replicas I and II. The whole process starting 
from the configuration given by Eq.(\ref{Ini2}) is repeated many times. 
The idea of DS is to look for averages 
$\langle D(t') \rangle$ 
of the Hamming distance 
\begin{equation}
D(t')= \frac{1}{L^2} 
\sum_{\bf R}^{L\times L} 
\big ( 1-\delta_{m^{\rm (I)}(t+t',{\bf R}),m^{\rm (II)}(t',{\bf R})} \big ) \,, 
\label{demspr2}
\end{equation}
where $D(0)=1/L^2$. In 
Fig.\ref{fig5} we show $\langle D(t') \rangle $ 
calculated as an average taken over the $1000$ steps 
of the steady-state. It is seen that DS technique 
confirms $\theta$-driven transition $\theta=\theta_{\rm crit}$ maintained 
previously by the direct inspection of $\langle \phi \rangle(\theta)$ 
dependence.  

Next, the system is studied via nonequilibrium 
relaxation method\cite{Ozeki2003} that is 
an efficient tool of analysis of equilibrium 
phase transitions. 
The numerical results are 
shown in Fig.\ref{fig6}. The initial configuration 
has been prepared in a such way that at the beginning, 
the factories are generated at random positions. 
Subsequently, the emissions have been left to increase 
freely without control. After the level 
$\sum_{\bf r}^{L\times L} m(t,{\bf r})=5000 L^2$ 
is reached, the sensors become active. The resulting 
relaxation dependence is then constructed by averaging over 
1000 repetitions of the whole relaxation process.
We found that relaxation strongly depends on how 
far $\theta$ is from $\theta_{\rm crit}$. Three distinct dynamical 
regimes have been identified when inspecting averaged 
relaxation process. In early  stage of development 
practically all sensors are active  
since $\langle \phi\rangle \sim 1$. 
Later, when content of the most of cells are still above 
$m_{\rm c}$, relatively fast exponential relaxation dominates. 
It corresponds to the purification through bounds. 
As it can be clearly seen, the processes at large times are 
well separated from the initial relaxation. The long-time 
process is associated with the positional adjustment 
(or self-organization) of the 
factories. 
\section{Conclusions}
The statistical properties of the multi-agent system 
including the environ\-men\-tal-econo\-mic interactions have 
been studied. It should be noted that 
at early stages of the work we were motivated by the particular 
hypothesis about implementation of feedback and direct generation 
of SOC regime. However, even preliminary runs have uncovered 
that not all the regimes conducted by feedback 
acquire SOC signatures as are the power-law distributions 
and power-law autocorrelation functions. With these facts in mind,
 we shift our focus to give an interpretation 
in terms of phase transitions ("unstable" 
in sense of Ref.\cite{Sornette}). 
Our simulation confirmed that only optimized 
feedback induces attractiveness of the critical state. 
In other words, the distributed feedback 
(as many other approaches) does not automatically 
guarantees the attractiveness of  critical point. 
The agents proposed here hold only rudimentary thinking 
and thus more intelligent entities should be considered 
in future work. A variety of problems can be formulated 
where the establishment of new factory 
is subjected to a place-search optimization strategy. 
With such modification the formation of highly correlated 
\emph{factory - sensorial agent} structures are expected.
One can extend the idea to models with 
crystal-like ordering and formation of the compact aggregates of factories - 
industrial complexes enhancing the effect of screening. 
An alternative representation of distributed feedback mechanism
may be the notion of individual energy taxation of factories which is proportional to 
the local contamination level.
Contrary to plethora of possible additional realistic 
entries omitted, the hypothesis posed by simulation 
is that the phase transition appears 
to be a generic statistical aspect of 
the groups of autonomous agents with conflicting 
interests. The main implication from 
the present model is the 
identification of power-law~(nearly catastrophic) 
pdf's of fluctuations of emission rates. 
We believe that focus 
to power-law distributions 
will open new perspectives of the 
environmen\-tal diagnostics.  

\ack 
The authors would like to express their 
thanks to Slovak Grant agency VEGA 
(grant no.1/2009/05) and 
grant agency APVT-51-052702 
for financial support.  


\begin{thebibliography}{}

\bibitem{Pearson1995} M.~Pearson, International Tax and Public Finance {\bf 2}, (1995) 357.

\bibitem{Speck1999} S.~Speck, Energy Policy {\bf 27} (1999) 659.

\bibitem{Ploeg1991} 
F.~Ploeg, A.~Zeeuw, System \& Control letters {\bf 17} (1991) 409.

\bibitem{Carraro1997}
C.~Carraro, M.~Galeotti, Energy Economics {\bf 19} (1997) 2.

\bibitem{Eftichios2005}
E.S.~Sartzetakis, P.~Tsigaris, J. of Regulatory Economics {\bf 28} (2005) 309.

\bibitem{Melby2000}
 P.~Melby, J.~Kaidel, N.~Weber, A.~H\"ubler, Phys. Rev. Lett. {\bf 84} (2000) 5991.

 \bibitem{Holyst2000}
  J.A.~Holyst, K.~Urbanowicz, Physica A {\bf 287} (2000) 587.

\bibitem{Youss1998}
  M.~Youssefmir, B.A.~Huberman, T.~Hogg, Computational Economics {\bf 12} (1998) 97;

\bibitem{Bradford1990}
J.~Bradford De Long, A.~Shleifer, L. H.~Summers, R. J.~Waldmann, 
Journal of Finance {\bf 45} (1990) 379.



\bibitem{Kadanoff}
L.P.~Kadanoff, Physics Today (March 1991) p.~9

\bibitem{Bak88}
   P.~Bak, C.~Tang, K.~Wiesenfeld, Phys. Rev. A {\bf 38} (1988) 364;

   P.~Bak, C.~Tang, K.~Wiesenfeld, Phys. Rev. Lett. {\bf 59} (1987) 381;

   C.~Tang, P.~Bak, Phys. Rev. Lett. {\bf 60} (1988) 2347.

 \bibitem{Sornette}
 D.~Sornete, J. Phys. I  France {\bf 2} (1992) 2065;

 D.~Sornette, A.~Johansen, I.~Dornic, J. Phys. I France {\bf 5} (1995) 325.
%
\bibitem{Tomita2001}
  Y.~Tomita, Y.~Okabe, Phys. Rev. Lett. {\bf 86} (2001) 572.

 \bibitem{HorvathGmitra2004}
  D.~Horv\'ath, M.~Gmitra, Z.~Kuscsik, Czech. J. Phys. {\bf 54} (2004) 921;

  D.~Horv\'ath, M.~Gmitra, Int. J. Mod. Phys. C {\bf 15} (2004) 1269;

  M.~Gmitra, D.~Horv\'ath, Int. J. Mod. Phys. C {\bf 16} (2005) 1943;

  D.~Horv\'ath, M.~Gmitra, Z.~Kuscsik, Physica A, {\bf 361} (2006) 589.    

 \bibitem{Clarke1994}
 H.R.~Clarke, W.J.Reed, J. of Economic Dynamics and Control {\bf 18} (1994) 991.

 \bibitem{Tsur1998}
 Y.~Tsur, A.~Zemel, J. of Economic Dynamics and Control {\bf 22} (1998) 967.

\bibitem{Axtell}
R.L. Axtell, C.J. Andrews, M.J. Small
 J. of Industrial Ecology {\bf 5} (2006) 10.



\bibitem{deCosta2006}
New Scientist magazine, (2. Febr. 2006),  29.

\bibitem{Dhariwal}
A.~Dhariwal, G.S.~Sukhatme, A.~Requicha, 
IEEE International Conference on Robotics and Automation, 
New Orleans, Lousiana, Apr.2004, 1436-1443.


\bibitem{Marin2000} 
M.~Mar\'{\i}n, V.~Rauch, A.~Rojas-Molina, C.S.~Lopez-Cajun, 
A.~Herrera, V.M.~Castano, Computational Materials Science 
{\bf 18} (2000) 132.

\bibitem{Cammann1992}
K.~Cammann, Sensors and Actuators B {\bf 6} (1992) 19.

\bibitem{Arthur1989} W.B.~Arthur, The Economic Journal, {\bf 99} (1989) 116.


\bibitem{Fujiwara2004} 
Y.~Fujiwara, H.~Aoyama, C.~Di Guilmi, W.~Souma, 
M.~Gallegati, Physica A, {\bf 334} (2004) 112. 


\bibitem{Kauffman69}
S.~Kauffman, J.Theor.Biol. {\bf 22} (1969) 437. 

H.E.~Stanley, D.~Stauffer, J.~Kertesz, 
H.J.~Herrmann, 
Phys. Rev. Lett. {\bf 59} (1987) 2376;

E.M.~Sousa Luz, M.P.~Almeida, U.M.S.~Costa, M.L.~Lyra, Physica 
A {\bf 282} (2000) 176.

\bibitem{Ozeki2003} Y.~Ozeki, K.~Kasono, N.~Ito, S.~Miyashita, Physica A, 
{\bf 321} (2003) 271.  

\end{thebibliography}
\end{document}